\documentclass[aps,twocolumn,floats,prl,nofootinbib,superscriptaddress,10pt]{revtex4-1}

\usepackage[utf8]{inputenc}

\usepackage{graphicx,amsmath,amsfonts,amssymb,slashed}
\usepackage{bbold,wasysym}

\usepackage{amsmath,amsfonts,bbm,euscript,hyperref}

\usepackage{xcolor}
\usepackage{color}
\usepackage{hyperref}

\hypersetup{colorlinks, citecolor=blue, linkcolor=red, urlcolor=blue}

\usepackage{graphicx,capt-of}

\newcounter{subfloat}

\newcounter{subfloat2}

\newcommand{\nco}{\newcommand}
\nco{\beq}{\begin{equation}} \nco{\eeq}{\end{equation}}
\nco{\beqa}{\begin{eqnarray}} \nco{\eeqa}{\end{eqnarray}}
\def\be{\begin{equation}}
\def\ee{\end{equation}}    
\def\baray{\begin{eqnarray}}
\def\earay{\end{eqnarray}}

\nco{\lra}{\leftrightarrow}

\nco{\sss}{\scriptscriptstyle} \nco{\dphi}{\varphi}
\nco{\lsim}{\mbox{\raisebox{-.6ex}{~$\stackrel{<}{\sim}$~}}}
\nco{\gsim}{\mbox{\raisebox{-.6ex}{~$\stackrel{>}{\sim}$~}}}

\def\IK{\relax{\rm I\kern-.20em K}}
\def\IM{\relax{\rm I\kern-.20em M}}

\def\lsim{\mbox{\raisebox{-.6ex}{~$\stackrel{<}{\sim}$~}}}
\def\gsim{\mbox{\raisebox{-.6ex}{~$\stackrel{>}{\sim}$~}}}
\def\sss{\scriptscriptstyle}

\def\ga{\mathrel{\raise.3ex\hbox{$>$\kern-.75em\lower1ex\hbox{$\sim$}}}}
\def\la{\mathrel{\raise.3ex\hbox{$<$\kern-.75em\lower1ex\hbox{$\sim$}}}}

\def\calP{\mathcal{P}}

\usepackage{amsmath}
\usepackage{lipsum}

\begin{document}

\title{Multi-messenger probes of inflationary fluctuations and primordial black holes
}

\author{ Caner \"Unal}
\affiliation{ CEICO, Institute of Physics of the Czech Academy of Sciences, Na Slovance 1999/2, 182 21 Prague, Czechia}

\author{ Ely D. Kovetz}
\affiliation{Department of Physics, Ben-Gurion University of the Negev, Be'er Sheva 84105, Israel}

\author{ Subodh P. Patil}
\affiliation{Niels Bohr International Academy, Niels Bohr Institute, Blegdamsvej 17, 2100 Copenhagen, Denmark}

\begin{abstract} 
Next generation cosmic microwave background spectral distortion and pulsar timing array experiments have the potential to probe primordial fluctuations at small scales with remarkable sensitivity. We demonstrate the potential of these probes to either detect signatures of primordial black holes (PBHs) sourced from primordial overdensities within the standard thermal history of the universe over a 13-decade mass range ${\cal O}(0.1-10^{12})M_\odot$, or constrain their existence to a negligible abundance.  Our conclusions are based only on global cosmological signals, and are robust under changes in i) the statistical properties of the primordial density fluctuations (whether Gaussian or non-Gaussian), ii) the merger and accretion history of the PBHs and assumptions about associated astrophysical processes, and iii) clustering statistics. Any positive detection of enhanced primordial fluctuations at small scales would have far-reaching implications from the content of dark matter to origin of BHs in the centers of galaxies, and to the field content of the inflation. On the other hand, their non-detection would also have important corollaries. For example, non-detection up to forecast sensitivities would tell us that PBHs larger than a fraction of a solar mass can constitute no more than a negligible fraction of dark matter. Moreover, non-detection will also rule out the scenario that PBHs generated by primordial overdensities could be the progenitors of super-massive black holes (SMBHs), of topical interest as there are only a few widely accepted proposals for the formation of SMBHs, an even more pressing question after the detection of active galactic nuclei over a billion solar masses at redshifts $z \geq 7$. Finally, non-detection sets the strongest bounds on the amplitude of small scale inflationary fluctuations for over 6 decades.
\end{abstract}

\maketitle

{\it Introduction.} Cosmic microwave background (CMB) spectral distortion \cite{Fixsen:1996nj, Fixsen:2009xn, Delabrouille:2019thj, Chluba:2019kpb} and pulsar timing array (PTA) \cite{Arzoumanian:2015liz,Lentati:2015qwp,Shannon:2015ect,Moore:2014lga} observations promise to reveal precious information about cosmic evolution from the inflationary epoch through to the late universe. A key aspect of CMB distortion and PTA observations is that they probe primordial fluctuations at comoving scales much smaller than those accessible to CMB anisotropy and large scale structure (LSS) surveys. In this study, we show that the two in tandem offer a multi-messenger window through which one can make definitive statements about  primordial black holes (PBHs) generated from enhanced curvature fluctuations. We extend previous studies~\cite{Bugaev:2010bb,Byrnes:2018txb,Sato-Polito:2019hws,Inomata:2018epa,Nakamadistortion,NakamainducedNG,refpbhinduced,Unal:2018yaa,Gow:2020bzo} and show how this mechanism can be either confirmed or ruled out as a primary channel to generate black holes in the mass range ${\cal O} (0.1-10^{12})  M_\odot$ that form a significant part of the cosmic energy budget within the standard thermal history  (See the bounds on smaller mass PBHs \cite{Wang:2019kaf}).

Current bounds allow for PBHs to constitute some fraction of the total dark matter density (for recent reviews, see Refs.~\cite{sasakireview, Carr:2020gox, Green:2020jor}, and a recent analysis, Ref.~\cite{DeLuca:2020qqa}). Confirmation of the existence of PBHs will be remarkable from the point of view of fundamental physics. Were they sourced from enhanced primordial curvature perturbations, their presence could reveal invaluable information about the relevant operators and field content that sourced primordial density fluctuations at far smaller scales than  probed by CMB anisotropy and LSS surveys. 

Meanwhile, non-detection of spectral distortion signals and stochastic gravitational wave backgrounds to forecast sensitivities of proposed experiments will set constraints on the dimensionless primordial power spectrum $\mathcal P_\zeta$ as strong as $\mathcal P_\zeta < {\cal O}(10^{-8})$ for $ 10^2 < k / {\rm Mpc^{-1}} < 10^4$, and $\mathcal P_\zeta < {\cal O}(10^{-5})$ for comoving scales $10^4 < k /  {\rm Mpc^{-1}} <10^7$. This would 
 imply that PBHs with masses ranging over 13 decades could make up only a negligible fraction of the cosmic energy density and rule them out as seeds for super-massive black holes (SMBHs), thereby allowing us to constrain one of the more plausible formation channels for SMBHs. Among various suggestions, the most widely accepted ways to produce SMBHs (see reviews \cite{smbhrev1,smbhrev2} for more details) are i) supercritical growth of tens of solar mass population III stars \cite{pop3smbh1,pop3smbh2,pop3smbh3}, ii) heavy seeds of ${\cal O}(10^4-10^6) M_\odot$ from the direct collapse of gas clouds \cite{dcbh1,dcbh2,dcbh3}, and iii) PBHs of ${\cal O}(10-10^3) M_\odot$ seeding SMBHs in the very early universe before recombination \cite{pbhsmbh1,pbhsmbh2,pbhsmbh3,pbhsmbhastro,Clesse:2015wea}. 

 We will show that these conclusions are robust under changes in i) the statistical properties of the fluctuations (whether Gaussian or non-Gaussian); ii) the PBH evolution history, namely their merger history and accretion rates, as well as any additional assumptions regarding associated astrophysical processes  (see e.g.\ Ref.~\cite{Poulin:2017bwe,Mack:2006gz, Ali-Haimoud:2016mbv,Raidal:2018bbj,Inman:2019wvr,Vaskonen:2019jpv,DeLuca:2020fpg,Serpico:2020ehh}); and iii) clustering effects (see e.g. Refs.~\cite{Chisholm:2005vm,Clesse:2016vqa,Clesse:2017bsw,Desjacques:2018wuu,Ballesteros:2018swv,Bringmann:2018mxj,Belotsky:2018wph,Suyama:2019cst,Atal:2020igj}).

PBH formation from primordial density perturbations requires them to be significantly amplified at scales smaller than those probed by CMB anisotropies. One immediate consequence of this amplification is that gravitational waves are also sourced by anisotropic stresses quadratic in the scalar density perturbations. After horizon re-entry, these large density perturbations induce a potentially detectable stochastic gravitational wave background (GWB) via this nonlinear process, sourced schematically by an interaction of the form $\zeta+\zeta \to h$ \cite{Mollerach:2003nq,Ananda:2006af,Baumann:2007zm}. If these density perturbations correspond to comoving scales between $ {\cal O}(10^3-10^8) \, {\rm Mpc^{-1}}$, then the resulting GWB falls within a frequency range that is potentially detectable via pulsar timing arrays. Similarly, large density fluctuations can potentially induce observable $\mu-$distortions of the CMB spectrum through acoustic dissipation in the pre-recombination photon-baryon plasma if these enhanced perturbations correspond to comoving scales ${\cal O}(1-10^5) \, {\rm Mpc^{-1}}$, as CMB photons cannot thermalize perfectly for the modes that re-enter the horizon after $z\sim10^6$  \cite{Chluba:2015bqa}. Therefore, depending on the re-entry scale, enhanced primordial density fluctuations leave {\it inevitable} imprints on global cosmological signals. These signals are unavoidable since they do not require any further assumptions other than the standard thermal history of the universe supplemented with amplified primordial density fluctuations that could produce PBHs.

{\it Primordial Fluctuations, Statistics and PBH Abundance.} We start with the relation connecting the mass of a PBH and the wavenumber of the  fluctuations \cite{refpbhinduced}
\beqa
M_{\rm PBH, 0} &=& {\cal A}(M) \, {\cal M} (M)  \; M_{\rm seed}\nonumber\\
 & \simeq & {\cal A} \, {\cal M} \, 20 \,  \gamma \,  \left(\frac{k}{10^6 \, {\rm Mpc}^{-1}} \right)^{-2} \, M_\odot,
\label{eqmassk}
\eeqa
where ${\cal A}$ and  ${\cal M}$ are factors indicating the amount of mass gain due to accretion and the amount of merger for the corresponding BH mass, discussed further later. $\gamma$ is the ratio of the PBH mass to the mass inside the causal horizon. Under certain assumptions for PBH formation during radiation domination, we can set $\gamma$ to be 0.8 \cite{Germani:2018jgr}. However, it can also be viewed as a phenomenological parameter that varies under differing assumptions for the collapse mechanism which we discuss below.

The initial fraction of causal horizons that turn into PBHs can be calculated in either the Press-Schechter or the theory of peaks formalism as
\beq
\beta =  \int_{\Delta_c}^\infty P(\Delta) \, d\Delta,
\label{eqformfrac}
\eeq
where $P(\Delta)$ indicates the probability distribution function of $\Delta$, the density contrast, neglecting the smearing window function which is irrelevant for narrow spectra \cite{densitywindow}.  In Ref.~\cite{Green:2004wb}, it was found that both Press-Schechter and peaks-theory calculations were in relatively good agreement. However, a more refined treatment in Ref.~\cite{youngmassfrac} found that there was up to an order of magnitude discrepancy between the two approaches, although this is dwarfed by the uncertainty in precise value of $\Delta_c$, the critical threshold value for the formation of PBH. Numerical studies have shown that there is no universal value for the critical threshold, and that it lies roughly within the range $0.3 \la \Delta_c \la 0.66$ \cite{sasakireview}, an entirely natural consequence of the fact that differing radial profiles and shapes of the overdensities lead to collapse at different threshold values \cite{exc, musco, bbks, youngmassfrac,Germani:2018jgr,haradath}.

The source of  density perturbations are primordial curvature perturbations, typically taken to be Gaussian (G). If, as widely accepted, these primordial perturbations are inflationary in origin, they can be enhanced by various interactions in multi-field scenarios (such as gauge-axion coupling, waterfall transitions) or experience quantum diffusion and non-single clock effects in an ultra-slow roll stage. In such cases the density fluctuations can become non-Gaussian (NG)~\cite{Garcia-Bellido:2016dkw,Cook:2011hg,Barnaby:2012xt,Barnaby:2011vw,Thorne:2017jft,Namba:2020kij,Ozsoy:2020kat,Ozsoy:2020ccy,Kawasaki:2019hvt,Domcke:2017fix,Ballesteros:2017fsr,Garcia-Bellido:2017mdw,Germani:2017bcs,Motohashi:2017kbs, localNG,Nakama:2017ohe,byrnes2,s3,pina,byrnes,ngtwo,ng1,ng2,ng3,ng33,ng4, ng5,ng6,ng7,ng8,ng9,tailchisqr,Lyth:2012yp,Bugaev:2013fya,Linde:2012bt,Ozsoy:2018flq,Atal:2018neu}. Moreover, if certain interactions are strong enough, the curvature perturbations obey chi-square ($\chi^2$) statistics \cite{Lyth:2012yp,Linde:2012bt,Bugaev:2013fya,Garcia-Bellido:2016dkw,refpbhinduced,Kawasaki:2019hvt,Ozsoy:2020kat,Ozsoy:2020ccy}. As an exotic case, the cube of the normal distribution (G$^3$) can also be considered for the fluctuations, although this is much harder to produce as it requires the presence of additional irrelevant operators suppressed by some UV scale.

To capture a wide range of local non-Gaussianities, we parameterize the curvature perturbation as 
$\zeta = h(\zeta_{\rm G})$,
where $\zeta_{\rm G}$ is a Gaussian field. The local ansatz that informs the templates typically used to search for primordial non-Gaussianity can be expressed, for example, as
\beq
\label{hG}
h(\zeta_{\rm G}) = \zeta_{\rm G} + \frac{3}{5}f_{\rm NL}\left(\zeta_{\rm G}^2 - \sigma^2\right) + \frac{9}{25}g_{\rm NL}\zeta_{\rm G}^3 + ...
\eeq
from which the non-Gaussian probability distribution function for $\zeta$ is straightforwardly calculated as
\beq
P_{\rm NG}(\zeta) \mathrm{d} \zeta=\sum_{i=1}^{n}\left|\frac{\mathrm{d} h_{i}^{-1}(\zeta)}{\mathrm{d} \zeta}\right| P_{\rm G}\left(h^{-1}\right) \mathrm{d} \zeta,
\eeq
where the summation is over all the branches after having inverted the polynomial, Eq.~(\ref{hG}). For curvature fluctuations obeying G, $\chi^2$ and G$^3$ distributions, the formation fractions are given by evaluating the integral in Eq.~\eqref{eqformfrac} for the density perturbations expressed in terms of $\zeta$ and the corresponding $\zeta_c$, and are given by \cite{Lyth:2012yp,Linde:2012bt,byrnes,Garcia-Bellido:2016dkw,Nakama:2017ohe,refpbhinduced}
\beqa
\beta_{\rm G} &=&  {\rm Erfc} \left( \, \zeta_c \middle/ \sqrt{2 \mathcal P_\zeta / {\cal K} } \, \right), \nonumber\\ \nonumber\\
\beta_{\chi^2} &=& {\rm Erfc} \left( \, \sqrt{\frac{1}{2} + \frac{\zeta_c}{\sqrt{2\mathcal P_\zeta /{\cal K}} }} \, \right), \nonumber\\ \nonumber\\
\beta_{\, {\rm G}^{3}} &=& {\rm Erfc} \left[ \, \left( \zeta_c \middle/ \sqrt{ 8\mathcal P_\zeta /  15{\cal K}}  \right)^{1/3} \right].
\label{eqdifferentstat}
\eeqa
Some explanation is due here. First, we note that PBH formation is a highly nonlinear process, and the resulting abundance is affected by the nonlinear nature of gravity. Even if the primordial curvature perturbations were perfectly Gaussian, the density contrast relates to the curvature perturbation during radiation domination as 
\beq
\Delta=- \frac{8}{9} \frac{e^{-5\zeta/2} \, \nabla^2 e^{\zeta/2}}{a^2H^2},
\eeq
and this renders the overdensity and PBH formation inherently nonlinear. It has been found that this nonlinearity makes PBH formation harder. Incorporating the leading effects of these non-linearities within the Press-Schechter and peak theory formalism, results in a difference of approximately a factor of 2 with respect to the standard threshold integral \cite{Kawasaki:2019mbl,DeLuca:2019qsy,Young:2019yug}, which sets the parameter ${\cal K}\!=\!2$ in  eq \eqref{eqdifferentstat}. The former analyses were done for Gaussian curvature fluctuations (see also Ref.~\cite{Harada:2015yda,Germani:2019zez}).  Including effects of {\it mild} non-Gaussianities, one expects further $\mathcal{O}(1)$ difference, although significant deviations could result for much stronger ones \cite{Kalaja:2019uju}. Moreover, mode coupling also makes it hard to preserve a strictly $\chi^2$ distribution at small enough scales \cite{Young:2014oea}, however these uncertainties are mitigated for narrowly peaked spectra as we consider shortly. In what follows, we pick $\zeta_c=0.5$.

We can relate $\beta$, the fraction of causal horizons that collapse into PBHs as a given density perturbation mode re-enters the horizon during radiation domination, to the current energy density in PBHs of mass $M$, using \cite{refpbhinduced} 
\beqa
\Omega_{\rm PBH,0} && (M, {\cal A}, {\cal M}) = \frac{\rho_{\rm PBH,0}(M, {\cal A}, {\cal M})}{\rho_{\rm crit,0}} \nonumber\\
&& \simeq 2 \cdot 10^8 \gamma^{1/2} {\cal A} \sqrt{\frac{{\cal A} \, {\cal M} \, M_\odot }{M}}  \beta \left( \frac{M}{{\cal A} \, {\cal M} \, M_\odot} \right),
\label{eqpbhamount2}
\eeqa
where ${\cal A}$ and ${\cal M}$ encapsulate the PBH mass growth via accretion and mergers, respectively (assuming the initial mass function shifts to higher mass via mergers and accretion by constant amounts ${\cal A},{\cal M} \geq 1$). 
It is obvious that both ${\cal A}$ and ${\cal M}$ are functions of $M$ (for example, heavier BHs accrete more and one generally expects ${\cal A}\propto M^{p}, \, p>0$). However, the functional  dependence has no important effect on our conclusions since our main aim is to estimate the PBH energy density fraction, $\beta$, at the formation time. This quantity is extremely sensitive to the amplitude of the perturbations, and a few orders of magnitude change can be easily recovered by a factor of a few change in the density perturbation amplitude.
Integrating over all masses, we get the total energy density in PBHs today, 
$\Omega^{\rm tot}_{\rm PBH,0}=\int \Omega_{\rm PBH,0} \;  d\, \ln M$. 
Neither mergers nor clustering properties modify the global cosmological signals discussed in this work, furthermore they do not change the energy budget in PBHs  if GW leakage is ignored during mergers. Hence, our conclusions remain intact for different merger rates and clustering statistics.   

We note that while a broad density spectrum can modify local observables such as the BH mass function, clustering effects and forthcoming astrophysical processes,  it will only enhance the global cosmological signals we focus on (see \cite{Clesse:2018ogk,Orlofsky:2016vbd,Unal:2018yaa,Pi:2020otn}). To be conservative, we choose the narrowest possible spectrum (for a fixed PBH abundance, increasing the width by a few orders of magnitude only yields a factor of a few decrease in the density perturbation amplitude).

Another primary  goal of this work is to probe PBHs as SMBH seeds. We can estimate the approximate current energy density in SMBHs as follows: There are roughly $10^{12}$ galaxies within our current Hubble volume  \cite{refgalaxynum}, of which approximately $10\%$ are large galaxies, most of which are thought to have a SMBH at their galactic cores. Assuming their  average mass is between  $10^{6} - 10^8 M_\odot$, and given that the total mass in the universe is about $3\cdot 10^{57} {\rm gr}$, we see that  $M^{\rm tot}_{\rm SMBH} \sim  10^{11} \cdot (10^{6} - 10^8) M_\odot  \sim 2 \cdot (10^{50}- 10^{52}) {\rm gr}$, so that $ 10^{-7} \la\Omega_{\rm SMBH} \la  10^{-5}$. Probing $\Omega_{\rm PBH}$ down to $10^{-10}$, as we show is possible with future observations, may lead to a detection or quite conservatively imply that there \textit{must} be a formation channel for SMBHs other than enhanced primordial fluctuations.

As the primordial power spectrum is measured to be approximately scale invariant at scales probed by CMB anisotropies, it must grow substantially at smaller scales in order to cross the critical threshold to produce PBHs. It is common to consider delta function spikes in the power spectrum (which then induce log-normal mass functions), however this growth in power is unphysical. In Ref.~\cite{Byrnes:2018txb}, a steepest growth index of $k^4$ was found in the context of canonical single field inflation. In a two-field setting, allowing for isocurvature interactions, a steepest growth index of  $\sim k^{5}$ was uncovered \cite{Palma:2020ejf, Fumagalli:2020adf}\footnote{We note that within the canonical single field regime, Refs.~\cite{Carrilho:2019oqg} found a $k^5\log k^2$ steepest growth requiring phases of ultra slow-roll inflation and a preceding $\eta = -1$ phase such that $\calP_\zeta$ grows beyond eight orders of magnitude. Restricting $\calP_\zeta \leq 1$ recovers the $k^4$ steepest growth (see however \cite{ Ozsoy:2019lyy} for the effects of transients). In the non-canonical context, Ref.~\cite{Cai:2018tuh} proposed a resonant mechanism with a varying $c_s$ resulting in sharp spikes in the power spectrum, but only if $\dot c_s/(H c_s) \gtrsim 20$, not possible within the effectively single field regime \cite{Achucarro:2012sm}, necessitating a multi-field analysis where the very concept of an adiabatic sound speed loses meaning and would presumably return an index consistent with \cite{Palma:2020ejf,Fumagalli:2020adf}.}.

\begin{figure}[h!]
   \hspace*{.2in}
\centering 
\includegraphics[width=0.42 \textwidth,angle=0,scale=1]{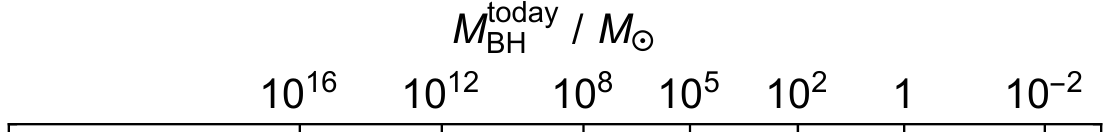}
  \vspace*{.07in}~\\
\includegraphics[width=0.49\textwidth,angle=0]{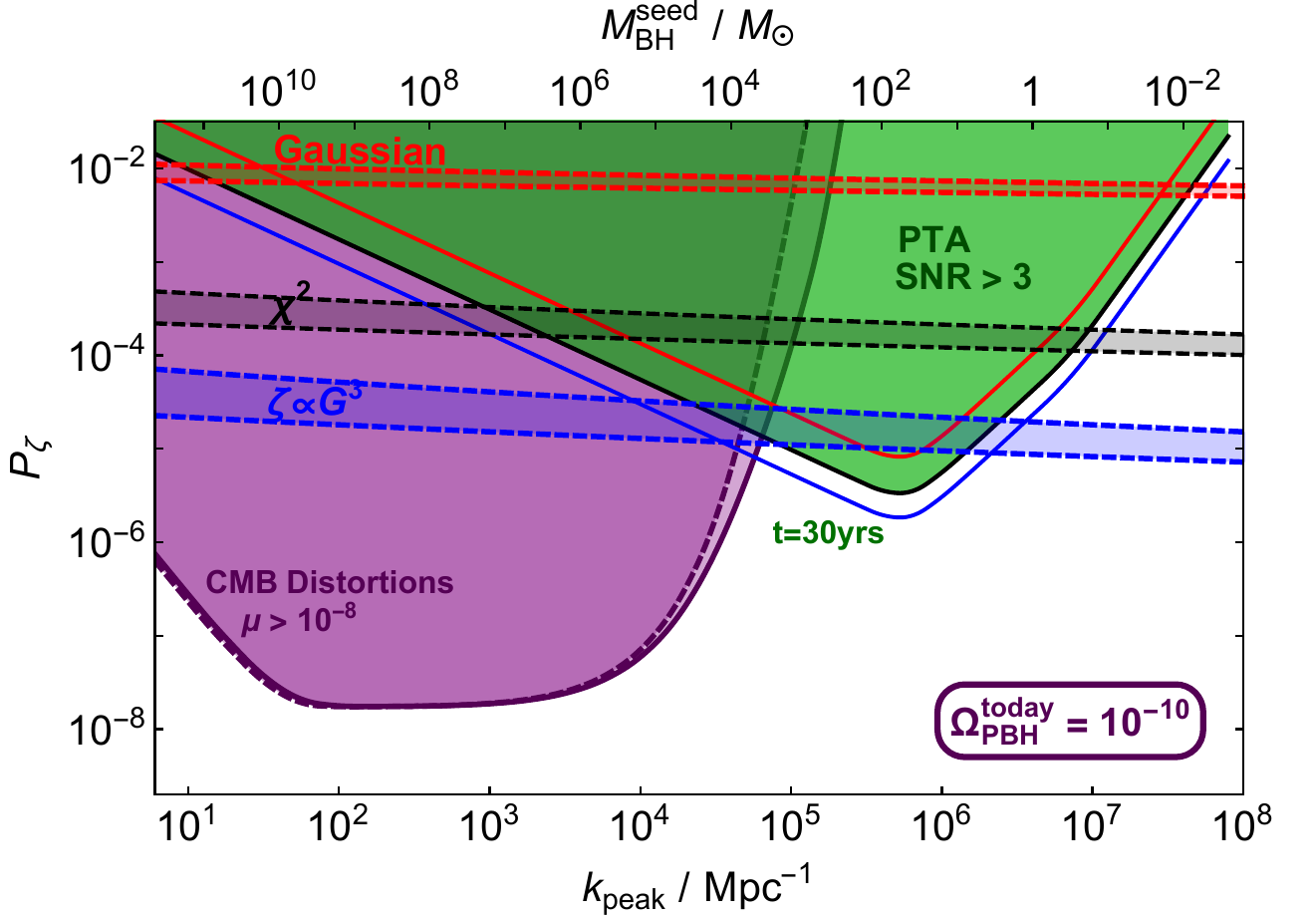}
\caption{Primordial perturbations that will be probed with CMB spectral distortions (SD) and PTA experiments. Solid (dashed) lines on small wavenumbers indicate SD bounds of $10^{-8}$ for $P_\zeta \propto k^4 (k^8)$. Solid lines at larger wavenumbers indicate the PTA bounds for different fluctuation statistics (red for Gaussian, green for $\chi^2$ and blue for $G^3$). Dashed horizontal bands indicate the amplitude of primordial perturbations which produces $\Omega_{{\rm PBH}}=10^{-10}$ given ${\cal A}=10^5$ (lower limit) and ${\cal A}=1$ (upper limit) accretion growth factors. On top, we show  the approximate current BH mass after evolution.
}
\label{figcombined}
\end{figure}

Therefore, we consider the power spectrum to satisfy some fastest growth  $\mathcal P_\zeta \propto k^n$ up to some scale $k_c$ where it must cut off. In order to be conservative (in the absence of any first principles derivation of a steepest growth index), we consider a range of values from $4 \leq n \leq 8$, where we take the latter as an extreme case to allow for multi-field effects. This leads to the following expression,
\beq
\mathcal P_\zeta = 2 \cdot 10^{-9} + n A^{\rm p} \left(\frac {k}{k_p}\right)^n \, \Theta(k_p-k),
\label{eqpz}
\eeq
where $\Theta$ is the Heaviside step function.
The primordial power spectrum amplitude $A^{\rm p}$ and its peak wavenumber $k_p$, given in Eq.~\eqref{eqpz}, are shown in dashed horizontal lines in Fig.~\ref{figcombined} for G, $\chi^2$ and G$^3$ statistical distributions to produce a given amount of PBHs, $\Omega_{\rm PBH} = 10^{-10}$. The upper horizontal dashed lines assume no accretion and mergers, while lower lines assume the PBH energy density grows by a factor of $10^5$ via accretion. This highlights two important points: i) An order unity change in the amplitude of the primordial power spectrum modifies the PBH amount by many more orders of magnitude (independent of the distribution), ii) NG fluctuations can significantly increase the efficiency of PBH production. In other words, the same PBH abundance can be produced by much smaller amplitude primordial density perturbations if they are non-Gaussian. A consequence of this is that both spectral distortions \cite{Chluba:2012gq,Chluba:2012we} and the secondarily produced GWs will also have a smaller amplitude for highly NG distributions that produce the same PBH abundance as Gaussian distributions \cite{refpbhinduced,Unal:2018yaa,Nakamadistortion,NakamainducedNG,Cai:2019elf,Cai:2018dig}. Therefore, the most conservative assumptions that allow for our conclusions to be robust when these assumptions are relaxed presume highly NG perturbations subject to reasonable presumptions about the steepest growth.

We now turn our attention to the signatures imprinted by enhanced primordial fluctuations onto two global cosmological observables: CMB spectral distortions and a secondarily sourced gravitational wave background.

{\it CMB Spectral $\mu$ Distortions.} The average (monopole) $\mu-$distortion in the CMB generated by acoustic dissipation of scalar perturbations in the photon-baryon plasma is given by \cite{Chluba:2015bqa,Nakama:2017ohe}
\begin{eqnarray} 
\!\!\!\!\!\!\!\! \!\!\!\! 
\langle \mu \rangle \simeq   2.3 \int_{k_0}^\infty \frac{dk}{k} \mathcal P_\zeta \left( k \right)  W(k),
\end{eqnarray} 
where the dimensionless power spectrum is convolved with a window function (where ${\hat k} = k \,/   \, {\rm Mpc}$ and ${\hat k}_0 =1$)
\begin{equation}
\!\!\! \!\!\!\! W(k) =  \left[ {\rm exp} \left( - \frac{\left[ \frac{\hat k}{1360} \right]^2}{1+ \left[ \frac{\hat k}{260} \right]^{0.3} + \frac{\hat k}{340}} \right) -  {\rm exp} \left( - \left[ \frac{\hat k}{32} \right]^2 \right) \right].
\end{equation}
$\mu-$distortions can only be efficiently produced after $z \sim 10^6$, as Compton scattering is highly efficient at thermalizing the photon-baryon plasma before this epoch. Therefore only those enhanced primordial curvature perturbations corresponding to comoving scales that re-enter the horizon in the window between $z \sim 10^6$ and recombination can produce non-negligible $\mu-$distortions. The horizon mass associated with these scales corresponds to PBHs in the range $10^2 - 10^{14} M_\odot$.

{\it Secondary/Induced Stochastic GW Background.}
Scalar curvature perturbations can source tensor perturbations at second order, and have been the subject of a number of studies, including Refs.~\cite{Mollerach:2003nq,Ananda:2006af,Baumann:2007zm} and more recently Refs.~\cite{Espinosa:2018eve,Kohri:2018awv,Tomikawa:2019tvi,DeLuca:2019ufz,Inomata:2019yww,Lu:2020diy,Domenech:2019quo}.
For enhanced curvature perturbations with a narrow width, the peak amplitude of the induced GW background is about the square of the peak amplitude of the scalar power spectrum. The same is also true up to an analytically calculable pre-factor for power-law scalar spectra \cite{Kohri:2018awv}. 

For IR modes, causality dictates that the energy density in GWs  scales with the third power of the frequency, i.e.\ $\Omega_{\rm GW}\propto f^3$ \cite{Unal:2018yaa,Cai:2019cdl,Yuan:2019wwo,Domenech:2020kqm}. In the UV regime, the GW background typically cuts-off sharply after $2k_{\rm peak}$ due to momentum conservation, as  the interaction that sources tensors from curvature perturbations is given by $\zeta+\zeta\to h$. Following the analytical results obtained for arbitrary scalar power spectra elaborated upon in Refs.~\cite{Kohri:2018awv,Pi:2020otn}, by defining $\tilde k=k/k_{\rm peak}$ and setting the peak location of the GW background as $\xi \simeq 1.2$, we have 
\beq
\frac{ \Omega_{\rm GW}^{\rm ind} h^2} {\Omega_{\rm rad,0} h^2  \cdot \mathcal P_\zeta^2} \simeq  \left( \Theta ({\tilde k}- \xi) \, F_{\rm IR} +  \Theta ({\tilde k}- \xi) \, F_{\rm UV}  \right) \Theta(2-{\tilde k}) \nonumber
\eeq
where 
$F_{\rm IR}=0.62 \, \left( {\tilde k}/ \xi \right)^3$
and 
$F_{\rm UV} =3 \, {\tilde k}^2 \left( 1-\frac{3 }{2} {\tilde k}^2 \right)^4
\times
 \left[ \frac{\pi^2}{4} \Theta(2-\sqrt{3}\, {\tilde k}) + \left( \frac{1}{2} \ln \bigg|1-\frac{4}{3 {\tilde k}^2} \bigg|  - \frac{1}{2- \frac{3 }{2}{\tilde k}^2}\right)^2 \right]$ 
 \\
correspond to wavenumbers smaller and larger than the peak scale of the induced GWs, respectively.

The statistical properties of the density fluctuations also have a marked influence on the amplitude and shape of the GW spectrum. In Ref.~\cite{Unal:2018yaa}, it was shown that due to a larger number of contractions resulting from higher symmetry factors and a larger number of non-vanishing diagrams, NG density fluctuations induce a larger GW amplitude --- by a factor of a few up to an order of magnitude depending on symmetry --- than those sourced by Gaussian density fluctuations of the same amplitude, and GW spectrum peaks at a higher frequency. (see also \cite{Ota:2020vfn,Yuan:2020iwf,Bartolo:2019zvb} for the further effects of NG on GW spectrum).

For temporal white noise, the energy density curve for PTA scales as the fifth power for frequencies larger than the inverse of the observation period  $f \cdot T_{\rm obs}>1$. We thus have $\Omega_{n} H_0^2= \frac{2 \pi^2}{3} f^3 \left( 12 \pi^2 f^2 {\cal S} \right) = 16 \,  \pi^4 \, T \, \lambda^2 \, f^5
$, where ${\cal S}= 2 \,  T \cdot \lambda^2$ as the noise power spectral density  \cite{Thrane:2013oya,Moore:2014lga} (see also Ref.~\cite{Schmitz:2020syl}). We assume typical values for all the parameters of the next generation PTA-SKA experiment, namely $ T=14$ days as the timing period and a $\lambda=30$ns  rms error in timing residuals. For frequencies $f \cdot T_{\rm obs}<1$, on the other hand, the noise energy density scales as 1/f 
 \cite{ptarealisticcurves,Janssen:2014dka,Feng:2020nyw}. In addition to this, in order to be conservative, we choose to include the effect of red noise, which was shown to be capable of decreasing sensitivity. We thus multiply the white noise result with an overall factor of 9 to get our noise spectrum (although red noise is typically more of an issue at smaller frequencies \cite{ptarealisticcurves,Mingarelli:2017fbe}). The energy density of the noise is thus expressed as 
\beq
\Omega_n (f) \, h^2  \simeq  1.5 \cdot 10^{30} 
\left( \left(\frac{{\rm sec}}{T_{\rm obs}} \right)^5 \, \left (\frac{1}{f\cdot T_{\rm obs}} \right) +   \left(  f \cdot {\rm sec} \right)^5   \right).
\eeq

Finally, we can express the signal-to-noise ratio (SNR) to reject the null hypothesis of no GW signal or estimate the signal amplitude (including the sample variance) as
\beqa\label{SNR}
&& {\rm SNR}_{\rm null}= \bigg[  2 \, T_{\rm obs} \left(\frac{{\cal N}^2} {2 \cdot 48} \right)  \int df \left(  \frac{\Omega_{\rm gw}}{\Omega_{n}}\right)^2 \bigg]^{1/2}, \nonumber \\
&& {\rm SNR}_{\rm det}= \bigg[  2 \, T_{\rm obs} \left(\frac{{\cal N}^2} {2 \cdot 48} \right)  \int df \left(  \frac{\Omega_{\rm gw}}{\Omega_{n}+\Omega_{\rm gw}}\right)^2 \bigg]^{1/2}
\eeqa
where ${\cal N}$ denotes the number of pulsars, $\Omega_{\rm noise}$ is the energy density of the noise, and $\Omega_{\rm gw}$ is that of the signal. The factor $2\!\cdot\! 48$ in the denominator results from angular averaging of Hellings \& Downs parameters \cite{Inomata:2018epa,ptacoeff,SGWBPTAstat,Chen:2019xse}.

{\it Results.} In Fig.~\ref{figcombined},  we show the combined bounds from CMB spectral distortions and PTA-SKA measurements on the primordial curvature fluctuations given different statistical properties (Gaussian in red, $\chi^2$ in black and $G^3$ in blue).  The purple solid(dashed) lines on  smaller wavenumbers indicate the  $10^{-8}$ $\mu$-distortion bounds,  typical for PIXIE-like experiment \cite{Kogut:2011xw,Kogut:2019vqh}, for $ \propto k^4 \; ( \propto k^8)$. The red, black and blue solid  lines at larger $k$ correspond to $3\sigma$ rejection of the null-hypothesis for Gaussian, $\chi^2$ and $G^3$ type perturbations, respectively. 
  We see that non-detection of signals by distortions and PTA-SKA concretely rules out any relevant $\gtrsim1M_\odot$ PBH abundance  independently from the fluctuation statistics. 
   
Any positive detection of a CMB $\mu-$distortion signal or a stochastic gravitational wave background from PTA observations will be uniquely exciting. If the signal can be verified to result from PBHs (and not other exotics), it will not only present a remarkable verification of the existence of PBHs as an ingredient of our universe, but  could also shed light on the relevant interactions, the shape of the potential or even the content of fields influencing the primordial density fluctuations at scales much smaller than hitherto probed. There is promise to be able to distinguish between the signatures of the PBH scenario considered here from other contributions to $\mu-$distortion or the GW background, especially using information from observables such as CMB and 21-cm anisotropies, the Lyman-$\alpha$ forest, etc.\ \cite{Ali-Haimoud:2016mbv,Poulin:2017bwe,Bernal:2017nec, Bernal:2017vvn, Murgia:2019duy,Carr:2020erq} (which are generally far more susceptible to astrophysical uncertainty). A detailed examination of this motivates future work.

\begin{figure}[h!]
\centering 
\includegraphics[width=0.49\textwidth,angle=0]{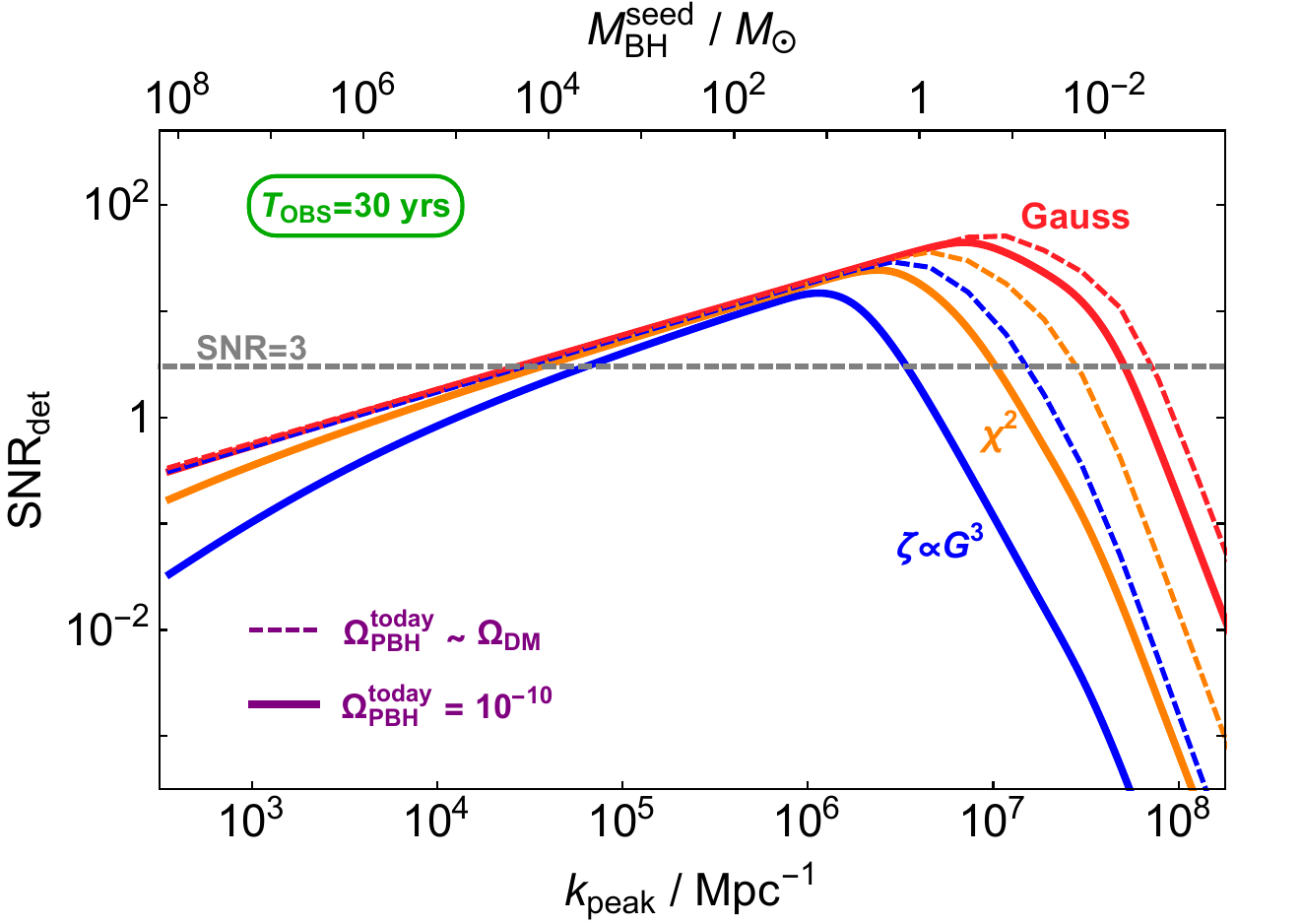}
\caption{ PTAs can detect PBH seeds for the mass range of ${\cal O} (0.1-10^5)M_\odot$ independent of the statistical properties of the fluctuations, be they Gaussian, $\chi^2$, or cubic non-Gaussian. Solid and dashed lines correspond to $\Omega_{\rm PBH}\sim 10^{-10}$ and $\Omega_{\rm DM}$ respectively.}
\label{figSMBHSNR}
\end{figure}

Assuming a detected signal is due to PBHs, we show in Fig.~\ref{figSMBHSNR} --- focusing specifically on the  capabilities of PTA-SKA --- our forecast sensitivity to the amplitude of the GW signal (see  ${\rm SNR}_{\rm det}$ in Eq.~\eqref{SNR}). The most interesting mass range for  massive SMBH seeds is ${\cal O}(1-10^5)M_\odot$. We see that with decades of observation, it will be possible to detect via the induced GW background a primordial signal which produces PBHs in the mass range $0.1-10^4M_\odot$ amounting to anything between an $\mathcal{O}(1)$ fraction of the total dark matter to only trace amounts of it, $\mathcal{O}(10^{-10})$, also for extreme statistical distributions. For higher mass seeds, the detection can be done by spectral distortions. 


{\it Conclusions.} To conclude, combined results from next-generation CMB spectral distortion and PTA experiments will be uniquely capable of detecting small scale primordial fluctuations which might result in primordial black holes formed from enhanced primordial fluctuations as a possible ingredient of the total matter in the Universe. We have studied the experimental sensitivity to their signatures  taking into account the induced GW shape, the statistical properties of density perturbations, pulsar number and red noise contamination, and applied conservative choices throughout in order to make our claims as robust as possible (we easily expect both constraints and measurement uncertainties to have bolder significance).

As we have shown, non-detection of spectral distortion or GWB signatures of PBHs from enhanced primordial fluctuations will have important consequences, namely that PBHs larger than a fraction of solar mass cannot constitute a relevant fraction of the cosmic energy budget.
This will  strongly rule out the scenario whereby PBHs originating from enhanced primordial fluctuations in the early Universe provided the seeds for SMBHs.

\section*{Acknowledgements}
We thank Christopher Berry, Christian Byrnes, Jens Chluba, Yi Feng, Juan Garcia-Bellido, Keisuke Inomata and Antonio Riotto for various discussions on PBHs, GWs and distortions, and especially Joseph Romano for his illuminating comments and explanations on PTAs.
C\"U dedicates this work to the memory of Metin Lokumcu, Ali Ulvi and Aysin B\"uy\"uknohut\c{c}u. C\"U is supported by European Structural and Investment Funds and the Czech Ministry of Education, Youth and Sports (Project CoGraDS - CZ.$02.1.01/0.0/0.0/15\_003/0000437$)  and thanks  Sabancı University, Ersin G\"o\u{g}\"{u}\c{s} and Yuki Kaneko for their hospitality. EDK is supported by a Faculty Fellowship from the Azrieli Foundation.

\end{document}